\documentclass[twoside,11pt]{article}

\usepackage{jmlr2e}

\usepackage[utf8]{inputenc}
\usepackage[T1]{fontenc}
\usepackage{newtxtext} 
\usepackage[scaled=.95]{cabin} 
\usepackage[varqu,varl]{inconsolata} 
\usepackage{amsmath}
\usepackage[varg]{newtxmath}
\usepackage{grffile}
\usepackage{longtable}
\usepackage{wrapfig}
\usepackage{rotating}
\usepackage[normalem]{ulem}
\usepackage{textcomp}
\usepackage{capt-of}
\usepackage[table]{xcolor}
\usepackage{subcaption}
\usepackage{hyperref}
\usepackage{microtype}

\ShortHeadings{Projection Predictive Inference for Generalized Linear and Additive Multilevel Models}{Catalina, B\"urkner and Vehtari}
\firstpageno{1}

\begin{document}

\setcitestyle{round}
\title{Projection Predictive Inference for Generalized Linear and Additive Multilevel Models}

\author{\name Alejandro Catalina \email alejandro.catalina@aalto.fi\\
    \addr Helsinki Institute for Information Technology, HIIT\\
    Department of Computer Science\\
    Aalto University\\
    02150, Espoo, Finland
    \AND
    \name Paul B\"urkner\thanks{Research done while in Aalto University as a postdoctoral researcher.} \email paul.buerkner@gmail.com\\
    \addr Cluster of Excellence SimTech\\
    University of Stuttgart\\
    70049, Stuttgart, Germany
    \AND
    \name Aki Vehtari \email aki.vehtari@aalto.fi\\
    \addr Helsinki Institute for Information Technology, HIIT\\
    Department of Computer Science\\
    Aalto University\\
    02150, Espoo, Finland}

\maketitle

\begin{abstract}%
    Projection predictive inference is a decision theoretic Bayesian approach that
    decouples model estimation from decision making. Given a reference model
    previously built including all variables present in the data, projection
    predictive inference projects its posterior onto a constrained space of a subset of variables. Variable selection is then performed by sequentially adding relevant variables until predictive performance is satisfactory. Previously,
    projection predictive inference has been demonstrated only for generalized
    linear models (GLMs) and Gaussian processes (GPs) where it showed superior
    performance to competing variable selection procedures. 
    In this work, we extend
    projection predictive inference to support variable and structure selection for
    generalized linear multilevel models (GLMMs) and generalized additive
    multilevel models (GAMMs). Our simulative and real-word experiments demonstrate that our method can drastically reduce the model complexity required to reach reference predictive performance and achieve good frequency properties.
\end{abstract}

\section{Introduction}
\label{sec:org6bf1d50}
Variable selection is an important aspect of statistical and predictive
modelling workflows, for example, when understanding a model's
predictions is important, or where there is a cost associated to collecting new data. 
From the perspective of predictive performance, one goal of
variable selection is to find the smallest subset of variables in a dataset
yielding comparable predictive performance to the full model containing all the
available variables. In this context, we assume that there might be variables
with true non-zero coefficients that we cannot properly detect due to scarce
data or the presence of a highly complex correlation structure.

In this paper, we substantially generalize projection predictive inference to
perform variable selection and model structure selection in generalized linear
multilevel models (GLMMs) \citep{mcculloch_GLMMs,Gelman_2013} and
generalized additive multilevel models (GAMMs)
\citep{Hastie_1986,Verbyla_1999}. Both types of models are widely used across
the quantitative sciences, for instance, in the social and political sciences
(e.g. poll or elections data whose measurements are organized in regions or
districts with multiple levels), or in the physical sciences (e.g.,
meteorological or medical data).

Projection predictive inference \citep{piironen_projective_2018} is a general
Bayesian decision theoretic framework that separates model estimation from
decision. Given a reference model on the basis of all variables, it aims at replacing its posterior
\(p(\lambda_{*} \mid {\cal D})\) with a constrained projection
\(q_{\bot}(\lambda)\). This projection is solved so that its predictions are as
close as possible to the reference model's predictions. The uncertainties in the reference
model related to the excluded model parts are also projected and thus partially
retained in the projection.

In the context of variable selection, one typically constrains the projection to
a smaller subset of variables where the excluded variables have their
coefficients fixed at zero. Then, the projection procedure sequentially projects 
the posterior onto an incremental subspace, until all the variables have entered 
the projection. At each step, the method selects the variable that most decreases 
the Kullback-Leibler (KL) divergence from the reference model's predictive
distribution to that of the projection model, a procedure known as forward search. 
This forms a \emph{solution path} for the variables into the projection. This approach 
has been shown to provide better performance than state-of-the-art competitors
\citep{piironen_comparison_2017,piironen_projective_2018,pavone_2019,piironen_projection_2016}.
\cite{piironen_comparison_2017} demonstrate that, when using the projection
approach, overfitting in the model space search is very small compared to
other stepwise procedures and that, even in huge model spaces, the
selected submodel has similar predictive performance to the
reference model.

\begin{figure}
    \centering
    \includegraphics[width=1.\textwidth]{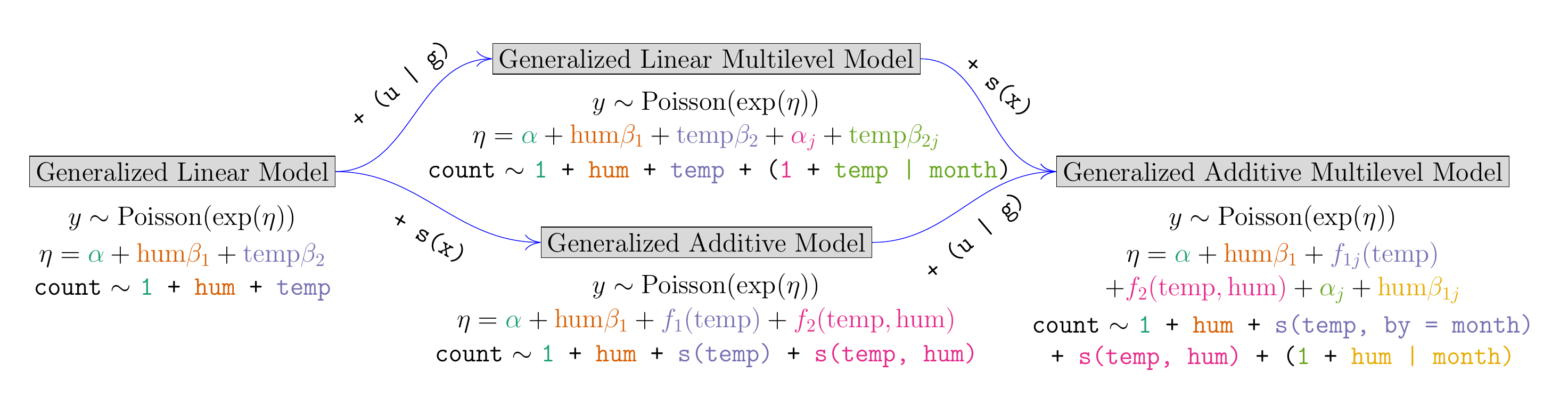}
    \caption{Different types of models  supported by \texttt{projpred} now (previously only generalized linear models) and their relationships. We showcase math and \texttt{R} \texttt{formula} correspondence in color-coded terms.}
    \label{fig:glm_gam_glmm_gamm}
\end{figure}

Previously, projection predictive inference has been used to perform
variable selection only in generalized linear models (GLMs)
\citep{piironen_projective_2018} and Gaussian processes (GPs)
\citep{piironen_projection_2016}.
However, the existing projection solutions do not directly translate to GLMMs or GAMMs
because, without further restrictions, the projection is not identifiable for
these models \citep{Bickel_1977}, that is, there is not a unique solution to
the projection. In this paper we extend the projection predictive
inference to GLMMs and GAMMs.
In Figure~\ref{fig:glm_gam_glmm_gamm} we showcase a broader picture of the different types of models that are now supported in our framework, starting from basic GLMs to very complex GAMMs. 
Along them, we show examples of equations for these kind of models and their correspondence to \texttt{R formula} syntax, which is an easy way of expressing complex models.

Specifically, our contributions include:

\begin{itemize}
\item Discussing the identifiability issue for projecting to GLMMs and GAMMs.
\item Extending projection predictive inference to support GLMMs and GAMMs by
performing a Laplace approximation to the marginal likelihood of the
projection.
\item Performing extensive simulations and real data experiments that validate the
working of our method.
\item Implementing our proposal in the open source \texttt{projpred} \texttt{R} package for
projection predictive inference \citep{projpred_package}.
\end{itemize}

\begin{figure}
    \centering
    \includegraphics[width=\textwidth]{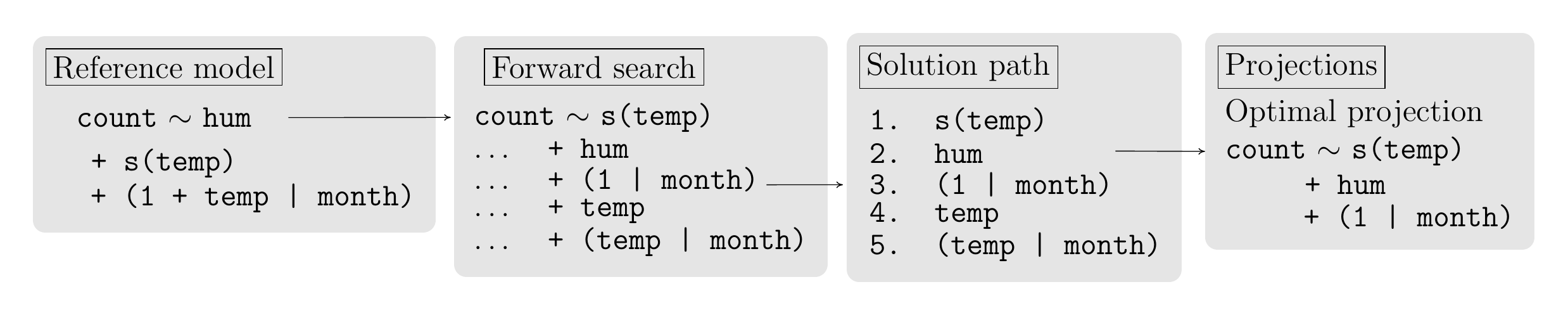}
    \caption{Projection predictive variable and structure selection workflow for an illustrative example on a \texttt{BikeSharing} data model.}
    \label{fig:projpred_workflow}
\end{figure}

To give an illustration of our developed method's application, consider the \texttt{BikeSharing} data. 
These data contain the hourly and daily count of rental bikes between 2011 and 2012 in London's capital bikeshare system with the corresponding weather and season information. 
The main variables included are \texttt{month}, \texttt{season}, \texttt{weather}, \texttt{temperature}, \texttt{humidity} and \texttt{windspeed}. 
In Figure~\ref{fig:projpred_workflow} we show the full projection predictive variable and structure selection for an illustrative reference model example for these data.
A priori we assume all continuous variables are relevant for predictions and they may interact with all categorical variables. 
For illustrative purposes we have built a simple model that contains different types of terms for only a subset of all the included variables, namely \texttt{hum, temp, month}.
We leave the remaining variables for a more in-depth analysis in the experiments in Section~\ref{sec:real_data_experiments}.
Our approach finds a projection model with simplified structure that provides optimal predictive performance with respect to the reference model. 

\section{Related methods}
\label{sec:org70ab7b4}
For GLMs, variable selection has been approached from different perspectives.
Some methods
\citep{breiman_garrote_1995,tibshirani_lasso_1996,fan_nonconcave_2001,Zou_2005,Candes_2007}
propose to deal with it by solving a penalized maximum likelihood formulation
that enforces sparse solutions, while at the same time trying to select a subset
of relevant variables (e.g. LASSO). These approaches suffer from confounding 
the estimation and selection of variables, often ending up
selecting fewer variables than truly relevant in the data, as in the case of
correlated covariates. For further information, see the comprehensive review by \citet{Hastie_2015}.
Similarly, \citet{Marra_2011} propose to add an additional penalty term to perform
variable selection in GAMs, with similar shrinkage capacity as ridge regression.
Another set of methods
\citep{George_1993,Raftery_1997,Ishwaran_2005,Johnson_2012,Carvalho_2010}
suggests imposing a sparsifying prior on the coefficients that favours sparse
solutions. Nonetheless, these priors do not actually produce sparse posteriors,
because every variable has a non-zero probability of being relevant. One can
obtain a truly sparse solution by selecting only those variables whose
probability of being relevant is above a certain threshold
\citep{Barbieri_2004,Ishwaran_2005,Narisetty_2014}, but this approach ignores the uncertainty in the variables below the threshold.

Reference models have been used before for tasks other than variable selection,
as in \citet{afrabandpey19:_makin_bayes_predic_model_inter}, where the authors
constrain the projection of a complex neural network to be interpretable (e.g.,
projecting onto decision trees). Closer to variable selection and related to our
approach, \citet{piironen_projection_2016} use projection
predictive inference and impose further constraints on the projection of a GP
reference model to perform variable selection, given the identifiability issue
of the direct projection.

While some alternative methods for variable selection in GLMMs and GAMMs exist,
they either only allow variable selection for population parameters but not
group parameters \citep{Groll_2012,Tutz_2012} or, when trying to use Bayes
factors \citep{Kass_1995}, are computationally infeasible due to combinatorial
explosion as soon as there are more than just a few variables. To the
best of our knowledge, there are no practically applicable competing methods
available in the literature, and so we only focus on the absolute performance of
our method.

\section{Projection Predictive Inference}
\label{sec:orgce68568}
\subsection{Formulation of the KL projection}
\label{sec:org151ebe8}
Because the domain of both models may be different, formulating the problem in
terms of minimizing a discrepancy measure between \(p \left( \lambda_{*} \mid
{\cal D} \right)\) and \(q_{\bot}(\lambda)\) does not make sense. Instead, we
minimize the KL divergence from the reference model's predictive distribution to
that of the constrained projection, which is not easy in its general form:
\begin{align}
\label{eq:kl_minimization}
\text{KL} &\left( p \left( \tilde{y} \mid {\cal D} \right) \parallel q \left( \tilde{y} \right) \right) \nonumber\\
& = \mathbb{E}_{\tilde{y}} \left( \log p \left( \tilde{y} \mid {\cal D} \right) - \log q \left( \tilde{y} \right) \right) \nonumber\\
& = - \mathbb{E}_{\tilde{y}} \left( \log q \left( \tilde{y} \right) \right) + \text{C} \nonumber\\
& = - \mathbb{E}_{\tilde{y}} \left( \log \mathbb{E}_{\lambda}\left( p \left( \tilde{y} \mid \lambda \right) \right) \right) + \text{C} \nonumber\\
& = - \mathbb{E}_{\lambda_{*}} \left( \mathbb{E}_{\tilde{y} \mid \lambda_{*}} \left( \log \mathbb{E}_{\lambda} \left( p \left( \tilde{y} \mid \lambda \right) \right) \right) \right) + \text{C},
\end{align}
where we have collapsed terms that don't depend on \(\lambda\) into \(C\). Here, the
expectations over \(\lambda_{*}, \tilde{y} \mid \lambda_{*}, \lambda\) are over the
posterior \(p \left( \lambda_{*} \mid {\cal D} \right)\), the posterior predictive
distribution \(p \left( \tilde{y} \mid \lambda_{*} \right)\), and the constrained
projection \(q_{\bot} \left( \lambda \right)\), respectively.

In practice, we approximate the KL minimization by changing the order of the
integration and optimization in \(\mathbb{E}_{\tilde{y} \mid \lambda_{*}} \left(
\log \mathbb{E}_{\lambda} \left( p \left( \tilde{y} \mid \lambda \right) \right)
\right)\). To make this feasible, \citet{Goutis_1998} propose to do the projection
draw-by-draw, where we find a direct mapping from a posterior draw of the
reference model to the projection's constrained space.
\citet{piironen_projective_2018} propose a further speedup by demonstrating that
it is possible to solve the projection employing only a small subset of posterior
draws or even representative points that can be found, for instance, by
clustering.

\subsection{Variable and structure selection}
\label{sec:org202b444}
A high level overview of the variable
selection procedure of projective predictive inference includes the following steps
\citep{piironen_projective_2018}:

\begin{enumerate}
  \item Cluster the draws of the reference model's posterior.
  \item Perform forward search to determine the ordering of the terms for the
    projection. At each step include the term that most decreases the KL
    divergence between the reference model's predictions and the projection's.
  \item Sequentially, compute the projections adding one term at a time.
\end{enumerate}

For a more robust variable selection, we perform a leave-one-out (LOO) cross
validation procedure through the model space. In this approach, we repeat the full
forward search \(N\) times by performing the selection with \(N-1\) data points and
leaving the remaining point as test point each time, resulting in \(N\) different
solution paths. Instead of running the procedure for every observation, 
\citet{vehtari_pareto_2015} show that we can achieve a similarly robust selection by 
running the procedure only on a carefully selected subset of points based on their 
estimated Pareto-\(\hat{k}\) diagnostic.

In GLMMs and GAMMs, we no longer perform variable but model structure selection, that is, select additive model components to which we refer to as {\it terms}. In this context, a {\it term} may refer to a single variable with a single coefficient, a
group level term, corresponding to all the coefficients of the group's levels,
or a smooth term, corresponding to all the coefficients associated with the
smooth basis functions. The structure selection involves the same steps as the variable selection only that variables are replaced by terms.

\subsection{Solving the projection for exponential family models}
\label{sec:org77bc3b6}
For GLMs with observation models in the exponential
family \citep{McCullagh_1989}, projecting a draw \(\lambda_{*}\) from the
reference model's posterior to the projection space \(\lambda_{\bot}\) in
Equation \eqref{eq:kl_minimization} coincides exactly with maximizing its
likelihood under the projection model. Given a new observation \(\tilde{y}_i\),
with expectation over the reference model
\(\mu_i^{*} = \mathbb{E}_{\tilde{y} \mid \lambda_{*}}(\tilde{y}_i)\), this
reduces to \citep{piironen_projective_2018}:
\begin{equation}
\label{eq:projection_exponential_family}
\lambda_{\bot} = \arg \max_{\lambda} \sum_{i=1}^N\mu_{i}^{*}\xi_i(\lambda) - B(\xi_i(\lambda)),
\end{equation}
which does not depend on the dispersion parameter \(\phi\), for some function \(B\)
of the natural parameters \(\xi_i(\lambda)\).


The above projection holds for observation models other than Gaussian or link
functions other than the identity, as long as they belong to the exponential
family. In these cases, there is no closed form solution for the projection 
parameters, and we run iteratively reweighted least squares (IRLS), where
at every iteration one computes a pseudo Gaussian transformation  of each log 
likelihood \({\cal L}_i\) as a second order Taylor series expansion 
centered at the projection's prediction \citep{McCullagh_1989,Gelman_2013}.


\section{Projection Predictive Inference for GLMMs and GAMMs}
\label{sec:org08a4ded}
\subsection{Generalized Linear Multilevel Models}
\label{sec:org2378051}
GLMMs \citep{mcculloch_GLMMs,Gelman_2013} jointly
estimate both \emph{global population} and \emph{group-specific} parameters.
This approach allows the model to \emph{partially pool} information across
groups, which is particularly useful for the estimates of groups with few data
points. In this case, we refer to multilevel structure as terms arising from the
levels of a categorical variable and their interactions with other variables.

Given a response variable \(y\) for a population design matrix \(X\) with group
design matrix \(Z\), we can write a GLMM as
\(y \sim \pi \left( g(\eta), \phi \right)\), where \(g(\cdot)\) is the inverse
link function of the generalized family \(\pi\), and \(\phi\) is its dispersion
parameter. The only difference to a GLM comes in the linear predictor
\(\eta = X\beta + Zu\), where \(\beta\) are the population parameters and we add
the group parameters \(u \sim p(u \mid \theta)\), which may depend on some hyper
parameters \(\theta\). The goal is to accurately estimate the model parameters
\((\beta, u, \phi, \theta)\).

\subsection{Generalized Additive Multilevel Models}
\label{sec:org6163466}
GAMMs \citep{Hastie_1986,Verbyla_1999} add
further complexity to GLMMs by introducing smooth terms, which are presented as
a linear combination of non-linear basis functions.

As for GLMMs, we can formulate the model as
\(y \sim \pi \left( g(\eta), \phi \right)\). In the case of generalized additive
models (GAMs) without multilevel structure, the predictor \(\eta\) can be
written as \(\eta = \sum_{j=1}^J f_j(X)\), where each \(f_j\) is a function of
the predictor matrix \(X\) (in practice, each \(f_j\) uses only a subset of
columns of \(X\)). These functions are usually represented via additive spline
basis expansion \(f(x) = \sum_{k=1}^K \gamma_kb_k(x)\) with B-splines \(b_k(x)\)
\citep{Eilers_1996}. To avoid overfitting, we can either penalize some summary
of the spline basis coefficients, or equivalently write it as a GLMM by
splitting up the evaluated spline basis function into an unpenalized null-space
(appended into \(X\)) and a penalized space (appended as group variables into
\(Z\)) where the prior on \(u\) serves the same purpose \citep{Wood_2017}.
Standard multilevel terms of GLMMs can be combined with non-linear smooth terms
of GAMs to form the even more powerful GAMM model class \citep{Wood_2017}.
However, as soon as smooth terms are added and translated to the GLMM framework,
the resulting \(Z\) matrix becomes much denser than in a standard GLMM, thus
further complicating inference.

\subsection{Solving the projection for GLMMs}
\label{sec:org1ab1c1f}
Without further constraints, even if its observation model belongs to the
exponential family, the projection \eqref{eq:kl_minimization} is not
identifiable for GLMMs \citep{Bickel_1977,lee_nelder_96,Gelman_2013}. This
means that, given the mean prediction of the reference model \(\mu_i^{*}\), there
is no unique solution for the parameters in the projection model fitted
to \(\mu_i^{*}\).

To make the model identifiable and solve the projection, we propose to further
restrict it by integrating out the group parameters \(u\). The resulting likelihood
can be written as:
\begin{align}
\label{eq:glmm_likelihood}
    {\cal L} & \left( y \mid \beta, u, \phi \right)  = p \left( y \mid \beta, \phi, \theta \right) \nonumber\\
             & = \int p \left( y \mid \beta, \phi, u \right) p \left( u \mid \theta \right) du \nonumber\\
             & = \prod_i^N\int p \left( y_i \mid \beta, \phi_i, u \right) p \left( u_i \mid \theta \right) du_i,
\end{align}
where \(u_i\) are the group parameters belonging to observation \(i\) and we have
assumed conditional independence between datapoints given the group parameters.
This integral cannot be evaluated in closed form. For simple models with a
single group one can numerically integrate the above expression (employing
Gaussian-Hermite quadrature) but this quickly becomes infeasible for
higher dimensional problems. Maximizing this likelihood with respect to the
model parameters, following Equation \eqref{eq:projection_exponential_family}, gives
the constrained projection.

There are many references in the literature that focus on practical approaches
to obtain suitable approximate maximum likelihood estimates for GLMMs
\citep{McCulloch_1997,lme4,Lee_2001,Lee_2006,ogden13,Booth_1999}. For our
purposes, it is essential to use an approximate solution that still provides a
good proxy for the KL divergence minimising solutions to Equation
\eqref{eq:kl_minimization}. Some of the methods cited above provide accurate and
reliable solutions but often at the expense of a higher computational
cost \citep[e.g.,][]{ogden13}.

In the statistics literature, one finds simpler approximations that would scale
better for our case, such as the well known Restricted Maximum Likelihood
Estimates (REML) \citep{Lee_2001,Lee_2006}. This approach obtains a solution
by dealing with the group parameters as \emph{fixed} data and appending them as an
extension to the population parameters. One would then solve an augmented
GLM.



\subsection{Laplace Approximation}
\label{sec:org5673980}
The REML approach does not provide a tractable approximation to the log marginal
likelihood obtained from Equation \eqref{eq:glmm_likelihood}, which takes the form
\begin{equation*}
\label{eq:3}
\log {\cal L} \left( \beta, u, \phi \right) = \sum_{i=1}^N \log \left\{ \int \exp \left( h_i \right) du_i \right\},
\end{equation*}
where \(h_i = \log p(y_i \mid g(\eta_i), \phi_i) + \log p(u_i \mid \theta)\) is
the (unnormalized) log joint density. The integrals in the above equation do not
exist in closed form except for models where \(p(u \mid \theta)\) is conjugate
to the likelihood \(\pi\). Even in those cases, if the dimensionality of $u$ is
not very small, the computation is still intractable.

As a general purpose solution, we consider a first-order Laplace approximation
to the integral \citep{ha2009maximum,barndoff1989,lme4}. We split up the
integration problem into sub-problems that are easier to solve. Given a value of
\(\theta\), we can find the conditional mode \(\tilde{u}(\theta)\) and conditional
estimate \(\tilde{\beta}(\theta)\) by solving the following optimization problem
\begin{equation*}
\begin{bmatrix}
\tilde{u}(\theta) \\
\tilde{\beta}(\theta)
\end{bmatrix} = \text{arg}\max_{u,\beta} h(u \mid y, \beta, \phi, \theta),
\end{equation*}
as the parameters that maximize the likelihood.

Usually, we express the conditional density on the \emph{deviance scale}:
\begin{equation*}
\label{eq:pirls}
\begin{bmatrix}
\tilde{u}(\theta) \\
\tilde{\beta}(\theta)
\end{bmatrix} = \text{arg}\min_{u,\beta} -2 h(u \mid y, \beta, \phi, \theta).
\end{equation*}

This optimization problem can be solved efficiently using Penalized Iteratively
Re-weighted Least Squares (PIRLS), as implemented in the popular \texttt{lme4}
\citep{lme4} package. At each iteration, PIRLS performs a Gauss-Newton iteration 
in the space of \(u\) and \(\beta\). See 
\citet{lme4} for more details.

The second order Taylor series expansion of \(-2\log h\) at \(\tilde{u}(\theta)\)
and \(\tilde{\beta}(\theta)\) provides the Laplace approximation to the \emph{profiled
deviance}. On the deviance scale, the Laplace approximation is a function of the
so-called \emph{discrepancy} measure and takes a sum of squares form:
\begin{equation*}
\label{eq:discrepancy}
d(u \mid y, \theta, \beta) = \left\| W^{1/2}(\mu) \left[ y - \mu(u, \theta, \beta) \right] \right\|^2 + \left\| u \right\|^2,
\end{equation*}
where \(\mu = g(\eta(u, \theta, \beta))\) is the inverse link transformation of
the latent predictor \(\eta\), and \(W\) is a diagonal matrix of weights. Optimizing
this function with respect to \(\theta\) provides the maximum likelihood estimates
of \(\beta, \phi\) by substituting \(\theta_{\text{ML}}\) into
\(\tilde{\beta}(\theta)\) and solving for \(\phi\) in \(h(u \mid y, \beta,
\theta_{\text{ML}}, \phi)\).
Importantly, optimizing the Laplace approximation is a problem in the space of
constrained \(\theta\), which is usually small and therefore easy to solve efficiently.

\subsection{Solving the projection for GAMMs}
\label{sec:orgb3c6429}
The identifiability issue that exists in GLMMs is further aggravated by having a
dense \(Z\) matrix in GAMMs, which makes the likelihood in Equation
\eqref{eq:glmm_likelihood} intractable to compute even in conjugate Gaussian models
with a single smooth term. This also happens in GAMs without any multilevel
structure.

In order to make these models identifiable, one has to 1) impose a quadratic
penalty on the coefficients of the basis functions \citep{Wood_2017}, which
also helps in avoiding overfitting, and 2) integrate out the group parameters and
group smooth terms. Solving the resulting maximum likelihood equations has similar issues as in the plain GLMM case.
Given that GAMMs can be represented as GLMMs, the same Laplace approximation is
commonly used to obtain maximum likelihood estimates in these models
\citep{Wood_2010,Wood_2017}.

\subsection{Computational cost}
\label{sec:org4b6cdc8}
The computational budget of our approach is composed of the following components
\begin{itemize}
\item Running PIRLS, i.e. solving the projection, for a given subset of terms.
\item Solving the same projection for a number draws.
\item Performing forward search to explore the model space.
\item Running LOO cross-validation for many data points.
\end{itemize}

\citet{piironen_projective_2018} demonstrate that a small number of
posterior draws is  sufficient to find a good solution path, which saves
a lot of computation during the forward search. Nonetheless, running PIRLS for
complex multilevel models is still expensive especially when done repeatedly for different posterior draws.


Further, in our approach, we reduce the number of models to explore in forward
search by considering only those that are sensible according to common
modelling practices for GLMMs \citep{Gelman_2006}. This means that we only consider a
model with a certain group parameter if its population parameter has already
entered the projection. Likewise, we only consider an interaction between two
variables when both of them have already entered the projection separately. This
saves the method from exploring many models that are not considered sensible in
the first place.
\section{Experiments}
\label{sec:org6e08ade}
We now turn to validating our method in both simulated experiments and real
world datasets.

\subsection{Simulations}
\label{sec:org7d43805}
\begin{table*}[pt]
\caption{\label{tab:generation_settings} Data generation process settings for the simulations.}
\centering
\begin{tabular}{lll}
\hline
Parameter & Description & Values\\
\hline
\(D\) & Number of variables & 5, 7, 10\\
\(V\) & Percentage of \(D\) as group parameters & 0.33, 0.67, 1.0\\
\(K\) & Number of grouping factors & 1, 2, 3\\
\(\rho\) & Correlation factor & 0.0, 0.33, 0.67, 0.9\\
\(L\) & Levels per grouping factor & 5\\
\(N\) & Number of observations & 300\\
\(\pi\) & Observation model & Gaussian, Bernoulli\\
\(s\) & Sparsity & 0.4\\
\hline
\end{tabular}
\end{table*}

We first validate our method by running projection predictive variable and structure selection as implemented in \texttt{projpred} \citep{projpred_package} on extensive simulations.
We systematically test simple and more complex models with both an increasing number of grouping factors and variables.
We also consider correlations between different coefficients per level within a grouping factor.
The complete settings of the simulations are shown in Table \ref{tab:generation_settings}.

To reduce external noise, we fix the number of observations to $300$, the number of levels in each grouping factor to $5$ and the sparsity to $0.4$ to make sure that some terms are irrelevant.
For each simulation condition, we run $25$ data realisations.

The complete data generation process, common for all observation models \(\pi\), is
given as
\begin{align*}
x_{id} & \sim \text{Normal}(0, 1)\qquad \beta_d  \sim \text{Normal}(\mu_{b,f}, \sigma^2_{b,f}) \\
z_d  & \sim \text{Bernoulli}(p=0.6)\qquad g_{ik} \sim \text{DiscreteUniform}(1, L) \\
\mu_{g_k} & \sim \text{MultivariateNormal}( \mu_g, \sigma^2_gI) \\
v_d & \sim \text{Bernoulli}(p=V) \\
u_{lk} & \sim \text{MultivariateNormal}(\mu_{g_k}, \Sigma_{\sigma^2_g, \rho}) \\
\eta_i & = \sum_{d=0}^{D} z_d\beta_dx_{id} + \sum_{k=1}^K\sum_{l=1}^L\sum_{d=0}^{D} v_dz_du_{lkd}x_{g_{ik} = l,d} \\
y_i & \sim \pi(g(\eta_i), \phi_i),
\end{align*}
for data points \(i = 1,\ldots,N\), variables \(d = 0,\ldots,D\), grouping factors
\(k = 1,\ldots,K\), inverse link function \(g\), and covariance matrix
\(\Sigma_{\sigma^2_g,\rho}\) with diagonal entries \(\sigma_g^2\) and off diagonal
elements \(\rho\sigma_g^2\). We collapse the intercept (\(d = 0\)) into
\(\beta\), \(u\), so that we have \(D + 1\) dimensions by appending a column of ones
to \(X\). We choose the identity link function.
%
We fix the mean and standard deviation hyper parameters for the intercept $\mu_f, \sigma^2_f$ to $0$, $20$ respectively, for the main terms $\mu_{b,f}, \sigma^2_{b,f}$ to $5$, $10$ respectively and for the group terms $\mu_g, \sigma^2_g$ to $0$, $5$ respectively.
We choose large values to avoid simulating practically undetectable terms.

We sample group terms following a two-step procedure:

\begin{enumerate}
\item We first sample \(K\) means for all grouping factors from a \(D + 1\) dimensional
multivariate normal distribution as
\begin{equation*}
 \mu_{g_k} \sim \text{MultivariateNormal}( \mu_g, \sigma^2_gI).
\end{equation*}
\item Then, for each level \(l\) and grouping factor \(k\), we sample from a
\(D + 1\) dimensional multivariate normal (with mean $\mu_{g_k}$) as
\begin{equation*}
 u_{lk} \sim \text{MultivariateNormal}(\mu_{g_k}, \Sigma_{\sigma^2_g, \rho}).
\end{equation*}
\end{enumerate}

\begin{figure*}[pt]
\centering
\includegraphics[width=.99\linewidth]{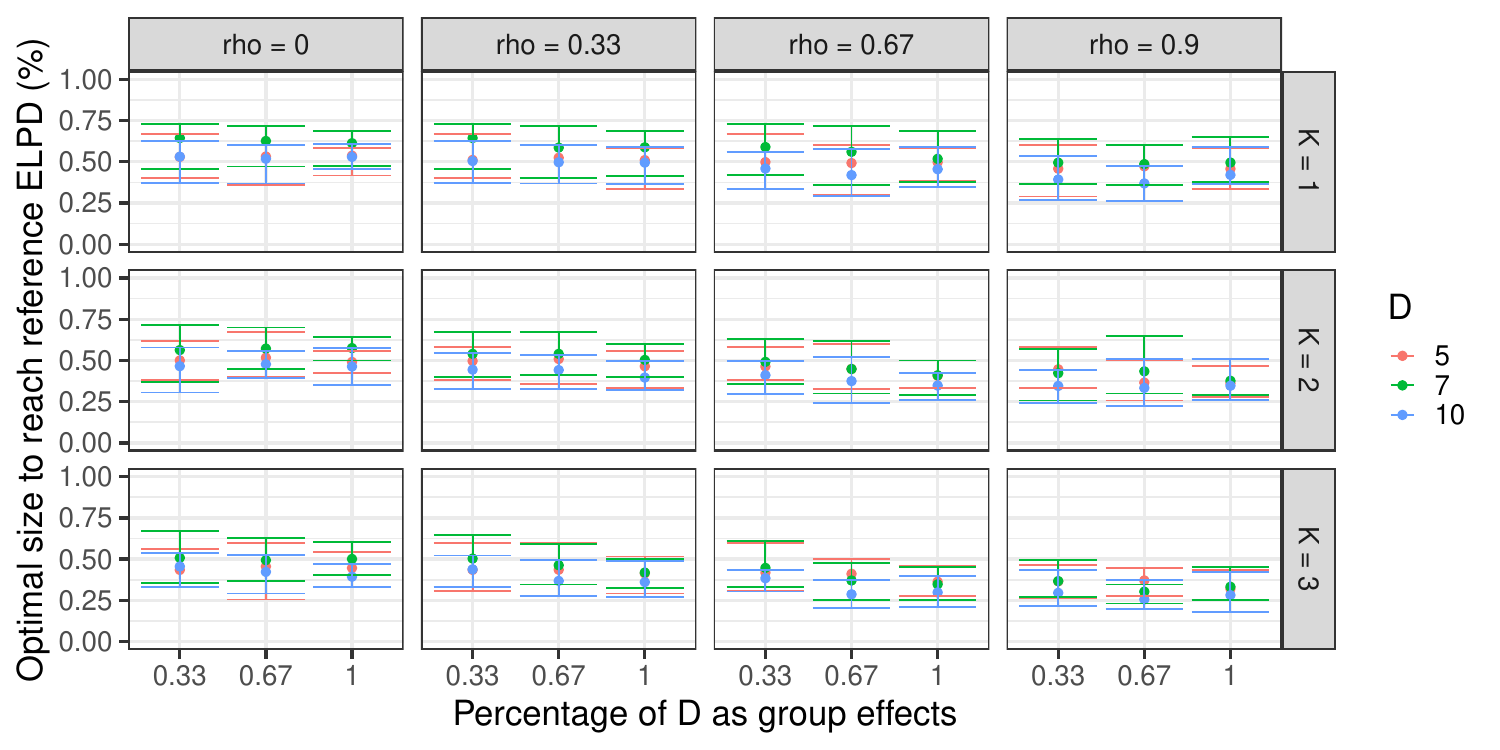}
\caption{\label{fig:gaussian_size_results}Optimal size for the projections to achieve the reference LOO performance (ELPD). Each column shows a different correlation factor. Each row shows a different number of grouping factors. We include 95\% uncertainty interval for the size of the projection.}
\end{figure*}

We first focus on studying the predictive performance of the projections. We
show the optimal size of the projection to reach the reference LOO Expected Log
Predictive Density (ELPD) performance for Gaussian simulated data in Figure
\ref{fig:gaussian_size_results}. We normalize this quantity by the total number
of possible terms for each model.

\paragraph{The projections achieve optimal predictive performance.} From the
predictive perspective, \texttt{projpred} aims at finding a projection with good
performance independently of the underlying true model. The results show that
the projection is able to discard on average 50\% of the terms without losing
performance. In the supplementary material we show these results for different
sparsity factors.

\paragraph{The projections are smaller with higher correlation.} Including correlated terms
into the projection would not improve its predictive performance, which results
in redundant components that can be discarded.

\paragraph{The projections are robust.} Even though the models become more and
more complex by adding more variables and a higher proportion of group terms,
the size of the optimal projection slightly decreases with respect to the total
number of terms. This also applies to the number of grouping factors.

\begin{figure*}[pt]
\centering
\includegraphics[width=.99\linewidth]{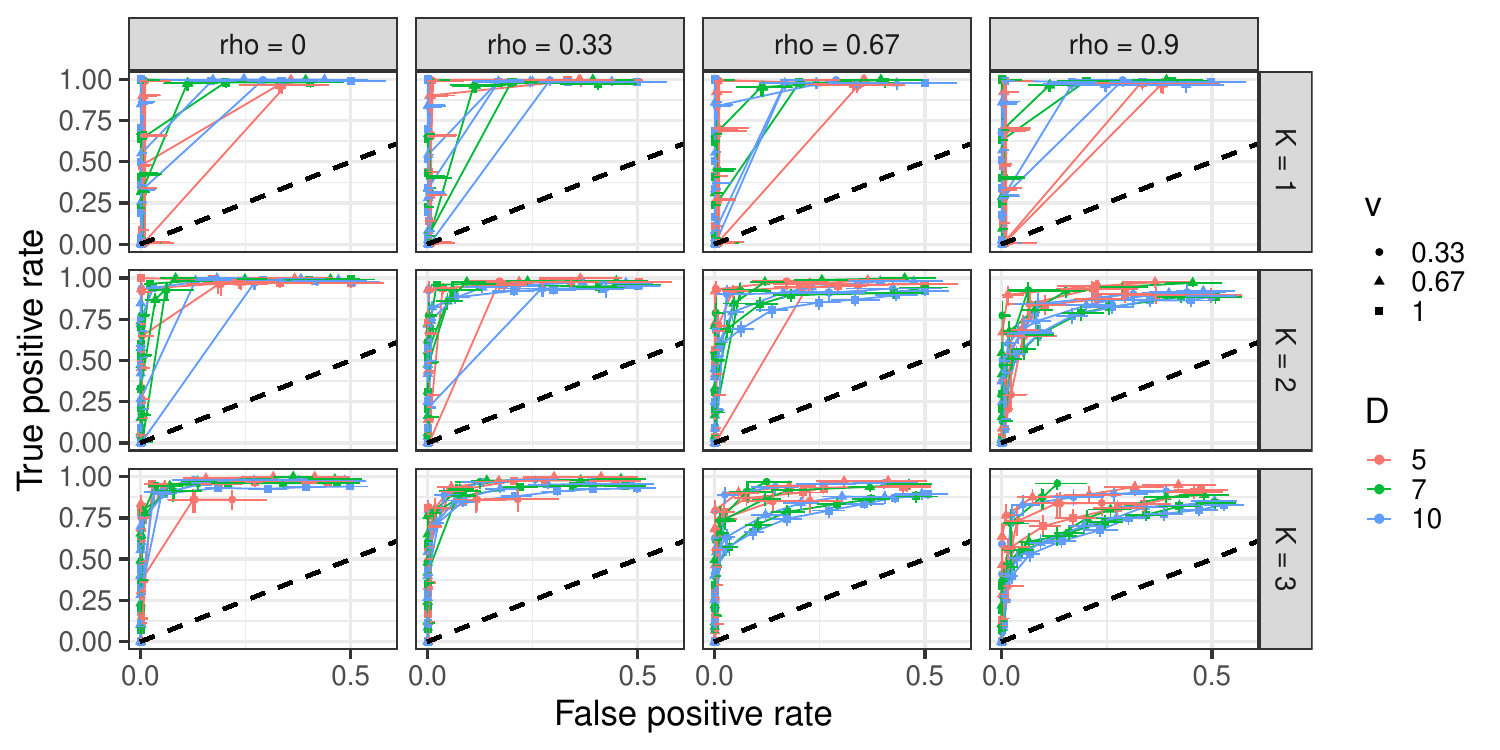}
\caption{\label{fig:gaussian_results}False against true positive rate in a Gaussian model for varying selection thresholds. Each column shows a different correlation factor. Each row shows a different number of grouping factors. We include 95\% uncertainty interval for both true and false positive rates. Chance selection in dashed black lines.}
\end{figure*}

We now turn our focus to study the selected group terms in the projections with
regard to their true relevance. Even though it is not \texttt{projpred}'s goal
to select all truly non-zero terms (but only all empirically relevant ones), it
is still useful to see how projection predictive selection coincides with the
known truth. We show these results for Gaussian data in Figure
\ref{fig:gaussian_results}. Importantly, even though in some cases our method
may miss relevant variables, it is able to find a projection with optimal
performance with just a subset of all variables and group coefficients, as shown
in Figure \ref{fig:gaussian_size_results}.

To compute true and false positive rates, we decide which terms are relevant by
looking at the ELPD improvement of each projection with respect to the previous
one. For a varying threshold \(t \in \left[ 0, 1 \right]\), we select as relevant
all terms whose projection's ELPD improvement is above the \(t\)th quantile of
all ELPD improvements. Then, we compare the selected terms against the  ground 
truth. For models with only a few terms, the discretization implied by the 
selection of terms results in some straight jumps in Figure \ref{fig:gaussian_results}.

\paragraph{The number of variables $D$ has a moderate effect.} By increasing the
number of variables we introduce more estimates in the model that in turn makes
the identification of the relevant terms harder. Then, as the dimension
increases, the true positive rate decreases while returning more false
positives.

\paragraph{The number of grouping factors $K$ is the most significant factor.}
Increasing the number of grouping factors in the data multiplies the number of
total parameters to estimate. This, in turn, dilutes the individual contribution
of each one and therefore makes its identification even more opaque. This is
reflected in the figure as we look at the bottom row, where the true positive
rate drops to above 75\% for 10 variables.

\paragraph{The percentage of group terms $V$ is important.} In the simulations,
we go to the extreme case of having all possible population terms vary across
all grouping factors, which, even though unrealistic, sets an interesting bar to
the performance of the method. In this extreme case, the false positive rate
reaches its maximum for all settings. Typically, though, only some terms vary
across grouping factors, and usually different terms vary between different
grouping factors.

\paragraph{High correlation $\rho$ induces more false positives.} Lastly, we
analyze the impact of the correlation. As the terms get more correlated, the
chance of selecting an irrelevant one as relevant gets higher, and therefore the
ratio of true and false positive rates worsens overall.

We show further simulations for Bernoulli and Poisson models in the
supplementary material. Furthermore, we also simulate more levels per grouping
factor and sparsity thresholds. Results of those simulations indicate similar
patterns and performance of our method to the results shown above.

\subsection{Real data experiments}
\label{sec:real_data_experiments}
We now turn to validating the performance of our method in real world datasets,
including a Bernoulli classification model and a Poisson count data model.

\subsubsection{Bernoulli classification model}
\label{sec:org1c840c3}
For the Bernoulli classification model we use the \texttt{VerbAgg} \citep{lme4}
dataset. This dataset includes item responses to a questionaire on verbal
aggression, used throughout \citet{de2004explanatory} to illustrate various
forms of item response models. It consists of 7584 responses of a total of 316
participants on 24 items. These items vary systematically in multiple aspects as
captured by three covariates whose parameters may vary over participants. For
the purpose of our study, we randomly draw \(50\) individuals and their
responses to increase the difficulty of the selection.

Following \texttt{R}'s \texttt{formula} syntax, we fit the reference model
\texttt{r2} \(\sim\) \texttt{btype + mode + situ + (btype + mode + situ | id)},
that includes \texttt{btype}, \texttt{situ} and \texttt{mode} as group
parameters varying over participants. The full reference model contains 7 terms,
counted as the simplest individual components of the model excluding the global
intercept, namely \texttt{btype}, \texttt{mode}, \texttt{situ}, \texttt{(btype |
  id)}, \texttt{(mode | id)}, \texttt{(situ | id)} and \texttt{(1 | id)}. 

We fit the reference model with a regularised horseshoe prior (with global scale
\(0.01\)) \citep{piironen_sparsity_2017,Carvalho_2010} with \texttt{rstanarm}
\citep{rstanarm} and run a leave-one-out (LOO) cross validated variable
selection procedure \citep{projpred_package,Vehtari_2016}.

\begin{figure}[pt]
\centering
\includegraphics[width=0.5\textwidth]{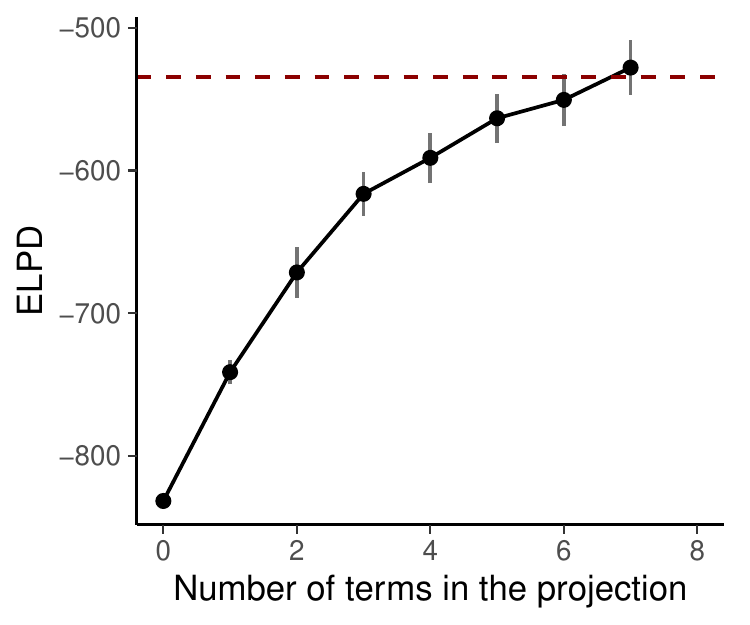}
\caption{\label{fig:elpd_report_verbagg}Summaries of incremental projections for \texttt{VerbAgg} dataset.}
\end{figure}

We show ELPD summaries and their standard error in Figure
\ref{fig:elpd_report_verbagg} for incremental projections. The red dash line
shows LOO performance for the full reference model.

The optimal projection threshold used by \texttt{projpred}
\citep{piironen_projective_2018} suggests including the first \(6\) terms,
resulting in the model \texttt{r2} \(\sim\) \texttt{btype + mode + situ + (btype
  + situ | id)}, effectively implying that almost all terms are relevant.
However, the standard error increases for the last two terms, which can be
explained by the correlations in the posterior, which may result in a slight
unidentifiability issue for those parameters. For more accurate projections one
could project more posterior draws or improve the robustness of the reference
model (e.g., with some PCA-like approach). The complete sequence of models
considered in the solution path is:

\begin{enumerate}
\itemsep0em
    \item \texttt{r2} $\sim$ \texttt{(1 | id)},
    \item \texttt{+ btype},
    \item \texttt{+ (btype | id)},
    \item \texttt{+ situ},
    \item \texttt{+ (situ | id)},
    \item \texttt{+ mode},
    \item \texttt{+ (mode | id)}.
\end{enumerate}

\subsubsection{Poisson GLMM model}
\label{sec:orge06671f}
For a Poisson count data model we use the \texttt{BikeSharing} dataset
\citep{bikesharing}. These data contain the hourly and daily count of rental
bikes between 2011 and 2012 in London's capital bikeshare system with the
corresponding weather and season information. We only use the daily averaged
dataset, with \(731\) observations for \(2\) years. It includes the following
variables: \emph{season}, \emph{month}, \emph{holiday}, \emph{weekday}, \emph{weather}, \emph{temp},
\emph{humidity}, \emph{windspeed} and \emph{count}. 

We build a knowingly overly complicated model that includes very correlated
group effects for different grouping factors, such as season, month or weather.
Our reference model is \texttt{count} \(\sim\)
\texttt{windspeed + temp + humidity + (windspeed + temp + humidity | month) + (windspeed + temp + humidity | weekday) + (windspeed + temp + humidity | weather) + (windspeed + temp + humidity | holiday) + (windspeed + temp + humidity | season)}.

We fit the reference model with a regularised horseshoe prior (with global
scale \(0.01\)) using \texttt{rstanarm} \citep{rstanarm} and provide it as an input to
\texttt{projpred}'s LOO cross-validated selection procedure.

\begin{figure}[pt]
\centering
\begin{subfigure}{0.49\textwidth}
\includegraphics{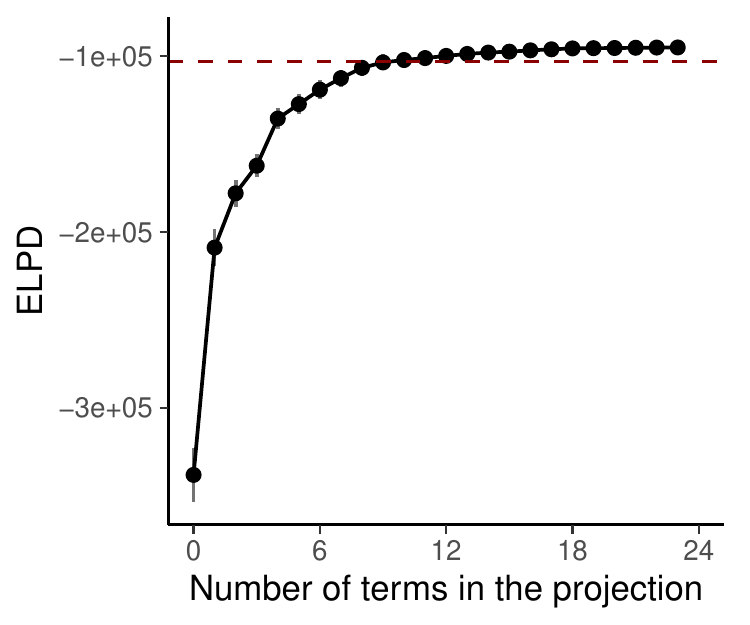}
\caption{\label{fig:elpd_report_bike}Summaries of incremental projections for GLMM model for \texttt{BikeSharing} dataset.}
\end{subfigure}
\begin{subfigure}{0.49\textwidth}
\includegraphics{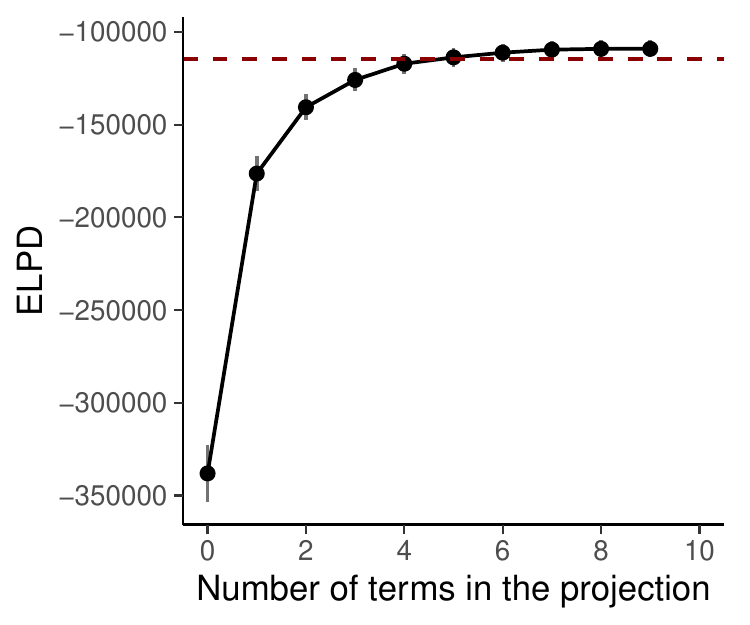}
\caption{\label{fig:elpd_report_bike_gamm}Summaries of incremental projections for GAMM model for \texttt{BikeSharing} dataset.}
\end{subfigure}
\caption{Summaries for \texttt{BikeSharing} models}
\end{figure}

We show ELPD summaries for each incremental projection in Figure
\ref{fig:elpd_report_bike}. The suggested optimal projection by \texttt{projpred}'s
heuristic method \citep{projpred_package} is \texttt{count} \(\sim\) \texttt{temp + humidity +
windspeed + (humidity + temp | month) + (1 | weather) + (1 | season)}, including
only \(8\) terms out of \(23\). The rest of the variables add only a marginal
improvement.

\subsubsection{Poisson GAMM model}
\label{sec:orge84f887}
We now used the same data as in the example above to build a simpler GAMM
reference model, for computational reasons, of the form \texttt{count} \(\sim\) \texttt{s(windspeed)
  + s(temp) + s(humidity) + (1 | month) + (1 | weekday) + (1 | weather) +
  (1 | holiday) + (1 | season)}, where we just added smooth terms for the main effects and group intercepts. 
Note that fitting GAMMs is usually more expensive and take longer time than plain GLMMs.

We make use of \texttt{rstanarm} again to build this model with a normal prior and
provide it as an input to \texttt{projpred}'s LOO-cross-validated selection.


We show ELPD summaries for each incremental projection in Figure
\ref{fig:elpd_report_bike_gamm}. The suggested optimal projection by
\texttt{projpred}'s heuristic method \citep{projpred_package} is \texttt{count}
\(\sim\) \texttt{s(temp) + s(humidity) + (1 | season) + s(windspeed) + (1 |
  month)}, including only \(5\) out of \(9\) terms.

\section{Discussion}
\label{sec:org153ee84}
In this work, we have extended projection predictive inference to variable and
structure selection on more complex classes of models, namely GLMMs and GAMMs.
For these models, the GLM projection solution cannot be directly translated as
it would result in unidentifiable models. For these model classes, combining
\texttt{projpred} with a Laplace approximation gives a good
approximation that not only enables accurate variable selection but also scales
well to larger number of variables and grouping factors.

We have validated our proposal by performing extensive simulations that test the
boundaries of our method in extreme settings. We also showed that our method
works well in real world scenarios with highly correlated grouping factors.

We leave the extension of our current framework to other models which do not
belong to the exponential family for future work. In such cases, the KL
minimization in Equation \eqref{eq:kl_minimization} does not coincide with
maximum likelihood estimates anymore.

\section*{Acknowledgements}

We would like to thank Akash Dhaka, Kunal Gosh, Charles Margossian and Topi Paananen for helpful comments and discussions. We also acknowledge the computational resources provided by the Aalto Science-IT project.

\newpage
\appendix

\section{Further results for more sparisty thresholds}
\label{sec:further-results-more}

In this section we provide more results for Gaussian simulated data with group
sparsity thresholds $s = \left\{ 0, 0.11, \ldots, 0.89 \right\}$. To isolate
the effect of the sparsity we fix the number of variables to $D = 5$ and $K = 1$
grouping factors. For a more complex setting, we allow all $D = 5$ variables to
vary within the grouping factor.

\begin{figure}[pt]
\centering
\begin{subfigure}{0.49\textwidth}
\includegraphics[width=0.99\linewidth]{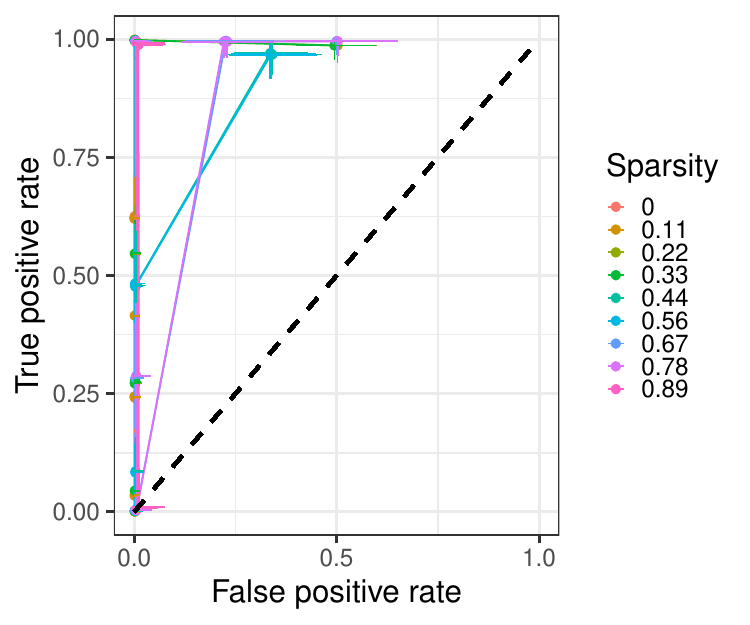}
\caption{\label{fig:gaussian_sparsity_results}False against true positive rate in a Gaussian model for varying sparsity thresholds. We include 95\% uncertainty interval for both true and false positive rates. Chance selection in dashed black lines.}
\end{subfigure}
\begin{subfigure}{0.49\textwidth}
\includegraphics[width=0.99\linewidth]{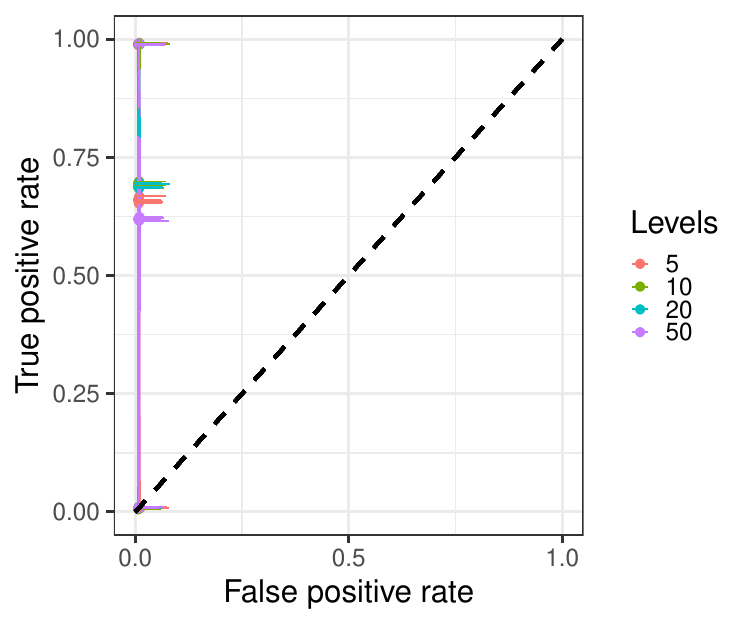}
\caption{\label{fig:gaussian_levels_results}False against true positive rate in a Gaussian model for varying number of levels. We include 95\% uncertainty interval for both true and false positive rates. Chance selection in dashed black lines.}
\end{subfigure}
\caption{Additional results for more sparsity factors and number of levels per grouping factor.}
\end{figure}

In Figure~\ref{fig:gaussian_sparsity_results} we show the ROC curve for this
experiment's results. To trace each sparsity's ROC curve we vary the selection
threshold for relevant group effects. As we did in the main text, we decide
which group effects are relevant by looking at the ELPD improvement of a
projection with respect to the previous one after including a given group
effect. For a selection threshold $t$, we select as relevant all group effects
whose projection's ELPD improvement is over the $t$th quantile of all ELPD
improvements. This gives a curve starting at $0$\% true positive rate when
$t = 0$, to $100$\% for $t = 1$ for every sparsity threshold.

Nonetheless, the sparsity threshold inherently sets the portion of actually
relevant group effects. As the sparsity increases, the number of relevant group
effects gets lower, and therefore the false positives rate can only increase as
we return more and more group effects as relevant. On the other hand, for low
sparsity thresholds we have the opposite behaviour, as there are many actually
relevant group effects, and therefore the false positive rate gets lower as we
select more group effects. The most difficult setting corresponds to medium
sparisty thresholds, and that is reflected in the figure with higher false
positive rates.

\section{Further results for more number of levels}
\label{sec:further-results-more-1}

In this case we focus on studying the effect of larger number of levels in each
grouping factor. We fix the number of variables $D = 5$ and $K = 1$ grouping
factor. We allow only $0.33$ portion of $D = 5$ total variables to be group
effects. We now vary the number of levels in $\left\{ 5, 10, 20, 50 \right\}$.
To avoid other factors to intervene, we simulate Gaussian observations.

We show ROC curves for all choices of levels in
Figure~\ref{fig:gaussian_levels_results}. We use the same mechanism to decide
relevant group effects and contrast them to the ground truth. In this case it's
clear that the number of levels is not inducing more false positives to be
returned, rather affecting only the true positives rate and how quickly the
method is able to get all of them right.

\section{Bernoulli simulations}
\label{sec:bern-poiss-simul}

Finally, we show optimal projection size and ROC curve results for Bernoulli
simulated data following the same structure as in the main text. In this case we
choose a \texttt{logit} link function. For non-Gaussian observation models there
is an added challenge as the projection is computing a linear Gaussian model as
an approximation to the non-Gaussian observation model. This increases the
computational budget of the method and sometimes results in unstable
projections, even though still resulting in good performance overall.

\begin{table}[pt]
\caption{\label{tab:hyper_parameters_bernoulli}Hyper parameters for the Bernoulli simulations. They are propagated to a constant vector of the corresponding value.}
\centering
\begin{tabular}{ll}
\hline
Parameter & Value \\
\hline
\(\mu_{f}\), \(\sigma^2_{f}\) & \(0\), \(4\) \\
\(\mu_{b,f}\), \(\sigma^2_{b,f}\) & \(0\), \(2\) \\
\(\mu_{g}\), \(\sigma^2_g\) & \(0\), \(3\) \\
\hline
\end{tabular}
\end{table}

On top of that, the hyper parameters we have used on the sampling procedure for
Bernoulli data imply smaller coefficients (see
Table~\ref{tab:hyper_parameters_bernoulli}). This is due the further complexity
added by the link function, that usually causes extreme values for large
effects.

\begin{figure*}[pt]
\centering
\includegraphics[width=.99\linewidth]{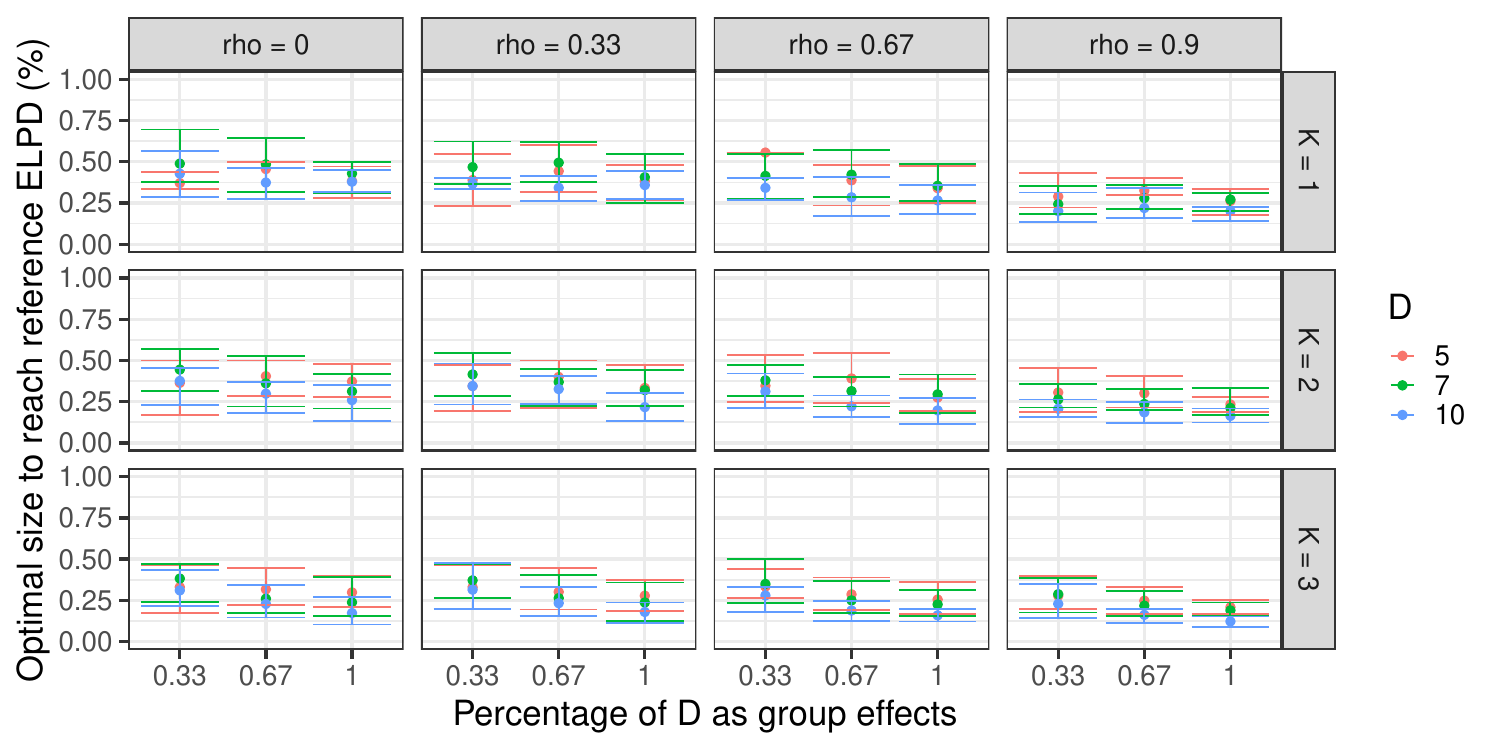}
\caption{\label{fig:bernoulli_size_results}Optimal size for the projections to achieve the reference LOO performance (ELPD). Each column shows a different correlation factor. Each row shows a different number of grouping factors. We include 95\% uncertainty interval for the size of the projection.}
\end{figure*}

We follow the same analysis structure as in the main text. We first focus on the
optimality of the projections and analyze the optimal projection size that the
method suggests to achieve the reference performance. We show these results in
Figure~\ref{fig:bernoulli_size_results}.

\paragraph{The projections achieve optimal predictive performance.} The results
show that the optimal projection is able to discard on average 60\% of the terms
while achieving the reference predictive performance. This proportion of relevant
terms is dependent on the sparsity threshold selected as analyzed in
Section~\ref{sec:further-results-more}. These results hold even for larger
number of grouping factors.

\paragraph{The projections are smaller with higher correlation.} In presence of
highly correlated variables the optimal projection is able to identify a smaller
subset of relevant variables that achieve the reference predictive projection.
As a result, the optimal size decreases as we increase the overall correlation.

\paragraph{The projections are robust.} Even as we increase the complexity of
the reference models, the method is able to find optimal projections every time.
The robustness of the results is indicated by the small confidence interval
around the mean estimates.

\begin{figure*}[pt]
\centering
\includegraphics[width=.99\linewidth]{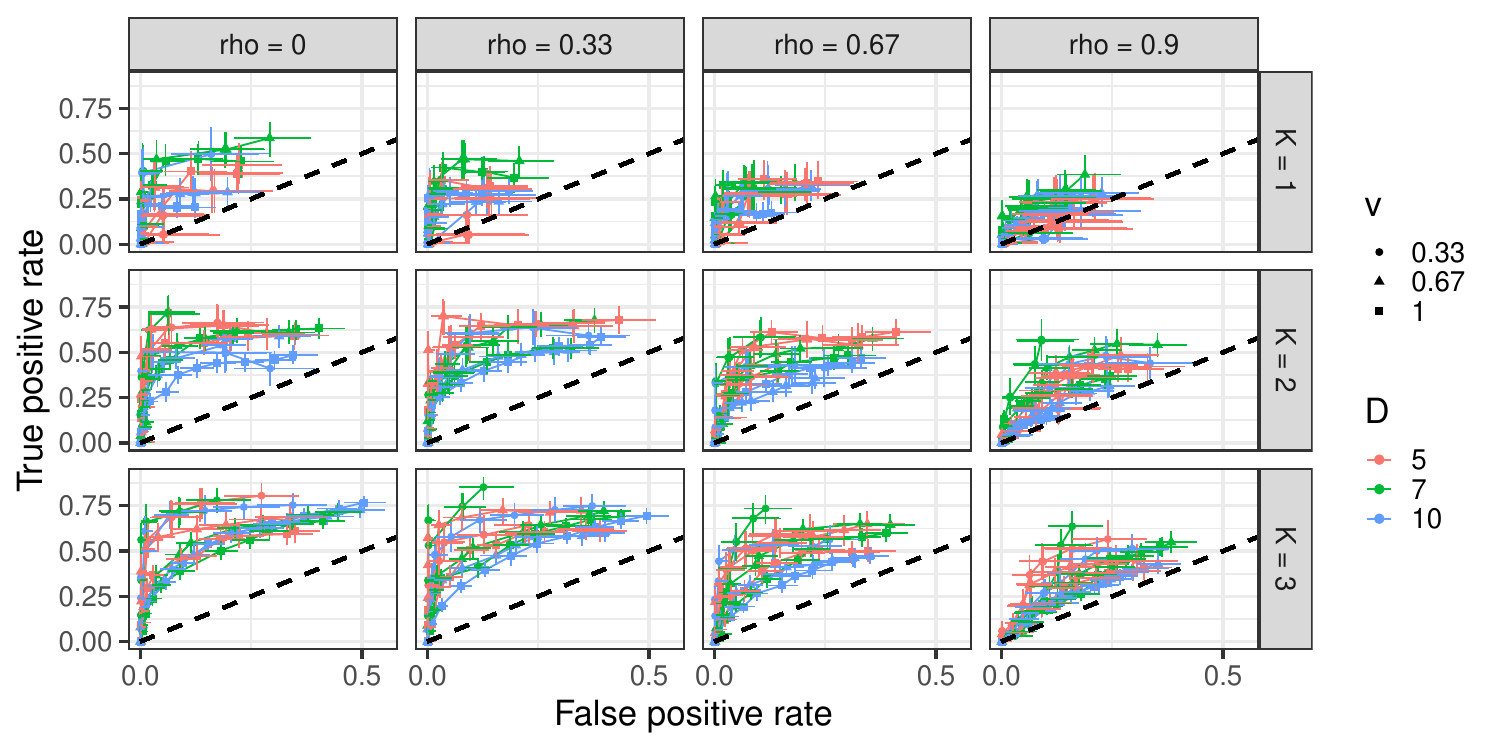}
\caption{\label{fig:bernoulli_results}False against true positive rate in a Bernoulli model for varying selection thresholds. Each column shows a different correlation factor. Each row shows a different number of grouping factors. We include 95\% uncertainty interval for both true and false positive rates. Chance selection in dashed black lines.}
\end{figure*}

Now, we analyze the performance of the method regarding the proper
identification of the truly relevant simulated terms. We show these results in
Figure~\ref{fig:bernoulli_results}. We employ the same method to compute true
and false positive rates for these experiments as we did in the main text for
Gaussian data. That is, for a varying threshold $t \in [0, \ldots, 1]$, we
select as relevant those effects that imply an ELPD improvement over the $t$th
quantile of all ELPD improvements after being included in the projection.

\paragraph{The number of variables $D$ has a moderate effect.} Increasing the
number of variables in the data has a direct effect on the complexity of the
reference model. This is also reflected in our results, as the ratio of true vs
false positive is better for smaller number of variables. Nonetheless, the
overall performance stays quite close.

\paragraph{The number of grouping factors $K$ is the most significant factor.}
Increasing the number of grouping factors in the data multiplies the number of
total parameters to estimate. As a result, the individual contribution of each
parameter is diluted, which makes its identification even more opaque. In the
figure we can clearly see how the ratio of true vs false positives rates gets
closer to random choice as we increase the number of grouping variables.

\paragraph{The percentage of group terms $V$ is important.} As we increase the
number of terms that vary within grouping factors we expect the method to
confound the relevance of each of them, as the variability of the data
increases. We can see, in turn, that the method returns more false positives for
higher proportions of group terms.

\paragraph{High correlation $\rho$ induces more false positives.} Another
important factor that induces more false positives is the correlation between
terms. This adds further complexity to the identification of truly relevant
terms and is clearly reflected in our results, as the ratio of true vs false
positive rates worsens for increasing correlation factors. Even in
the extreme unrealistic case of $0.9$ correlation, the method still identifies
around half of the truly relevant effects.

In general, the method does a good job at obtaining optimal projections with the
smallest possible subset of relevant terms that still achieves the reference
performance.

\bibliographystyle{apalike}
\bibliography{biblio}
\end{document}